\documentclass[preprint,12pt]{elsarticle}

\makeatletter
\def\ps@pprintTitle{%
 \let\@oddhead\@empty
 \let\@evenhead\@empty
 \def\@oddfoot{}%
 \let\@evenfoot\@oddfoot}
\makeatother

\usepackage{amssymb}
\usepackage{float}
\usepackage{subfigure}
\usepackage{amsmath}

\usepackage{graphicx,subfigmat,etoolbox,amssymb,float}

\begin{document}

\begin{frontmatter}



\title{Noise Measurement of a Wind Turbine using Thick Blades with Blunt Trailing Edge}

 \author[label2]{Weicheng Xue}

 \author[label2]{Bing Yang}
 \affiliation[label2]{organization={Institute of Engineering Thermophysics, Chinese Academy of Sciences},
             city={Beijing},
             postcode={100083},
             country={China}}             


\begin{abstract}
The noise generated by wind turbines can potentially cause significant harm to the ecological environment and the living conditions of residents. Therefore, a proper assessment of wind turbine noise is crucial. The IEC 61400-11 standard provides standardized guidelines for measuring turbine noise, facilitating the comparison of noise characteristics among different wind turbine models. This work aims to conduct a comprehensive noise measurement of a 100kW wind turbine using thick blades with blunt trailing edge, which differs from the typical turbines studied previously. The work takes into account the unique design and dynamic characteristics of small-scale wind turbines and adjusts the measurement accordingly, with deviations from the IEC standards will be explicitly addressed.
\end{abstract}




\end{frontmatter}


\section{Introduction}

The noise generated by wind turbines potentially has a significant impact on local residents' quality of life~\cite{nazir2020potential}. Therefore, a rational assessment of wind turbine noise is necessary. The IEC 61400-11 standard provides standardized guidelines for noise measurement of wind turbines~\cite{international2012iec}, enabling the comparison of noise characteristics among different wind turbine models. Previous studies conducted by King et al.~\cite{king2012assessing} measured four 2 MW wind turbines in Ireland continuously for 10 weeks, following the Irish regulations for measurement and calculation. They determined the critical wind speed for wind turbine operation, which is influenced by the turbine model, background noise level, and measurement location. Furthermore, the critical wind speed differs between daytime and nighttime. Migliore et al.~\cite{migliore2004acoustic} conducted a systematic noise assessment of eight small-scale wind turbines with power ranging from 400W to 100kW at NREL. They pointed out that the acoustic characteristics of wind turbines vary significantly even at the same wind speed due to climatic factors, and the variability of background noise severely affects the accuracy of noise assessments. Martens et al.~\cite{martens2020evaluation} presented five extensive measurement campaigns acquiring and analyzing the meteorological, acoustic and turbine-specific data at different locations in northern Germany, which lasted in total over 13 month. In order to obtain reliable measurements of the wind turbine noise, Gallo et al.~\cite{gallo2016procedure} presented a work to estimate the immission and the residual noise components measured nearby a wind farm, which allows the evaluation of the noise impact produced by operational wind farms. 

The distribution of wind resources in China~\cite{liu2019wind} differs from that of Western countries~\cite{gunturu2012characterization,decastro2019overview}, and it is necessary to develop wind turbine noise measurement standards suitable for the local conditions, especially when considering that the area of China is very large and the distribution of wind resources is very complicated. However, the IEC 61400-11 standard can still serve as a general method for noise measurement, facilitating the comparison and communication among different wind turbine noise assessments. Bo et al.~\cite{bo2011measurement} developed an IEC 61400-11 program for wind turbine noise measurement using the LabVIEW platform and summarized the advantages of real-time measurements using a virtual platform.

To develop wind energy resources in low-wind-speed and typhoon-prone areas, the Institute of Engineering Thermophysics of the Chinese Academy of Sciences designed a 100kW small wind turbine in Zhangbei County, Hebei Province, with autonomously designed thick blades and blunt trailing edges. This work focuses on the standardized noise measurement of this small-scale wind turbine using thick blades with blunt trailing edges~\cite{li2014large,li2016new,li2017experimental,li2020airfoil}. The noise characteristics of this small-scale wind turbine using thick blades with blunt trailing edges, are different from those of turbines using thin blades with sharp trailing edges. It should be noted that considering the different designs and dynamic characteristics of small-scale wind turbines compared to large-scale wind turbines, the noise measurement standards have been appropriately adjusted for this small-scale wind turbine. The procedures in which the measurements do not fully comply with the IEC standards will be explicitly addressed.

\section{Characteristics of the Wind Turbine}

\subsection{Blade Characteristics}

The blades of the 100kW wind turbine were designed by the Institute of Engineering Thermophysics of the Chinese Academy of Sciences, with the remaining components being completed by Shenyang Huarun Wind Power Co., Ltd. The blade design parameters were referenced from the relevant parameters provided by Huarun for the 100kW wind turbine and were designed using the NREL published blade design code HARP\_opt~\cite{sale2010preliminary}. The blade profile at the maximum chord length adopts the blunt trailing edge airfoil CAS-W2-450~\cite{wu2020uncertainty}, while the blade tip uses the NACA0018 airfoil~\cite{nakano2007experimental}, and all other airfoils adopt the DU series~\cite{timmer2003summary}.

\subsection{Main Technical Parameters}

The wind turbine is a horizontal-axis wind turbine, with a downwind rotor configuration. The hub height is 26.2 m, and the horizontal distance from the wind wheel center to the tower axis is 0.9 m. The rotor radius is 10.092 m, and it adopts pitch control and variable speed operation. The tower is of a cylindrical structure.

\section{Natural Environment}

The wind turbine is located in a large wind farm in Zhangbei County, Zhangjiakou City. To the south of this wind turbine, there is a large photovoltaic power generation test base, while to the north, there is a road several hundred meters away. Three abandoned small wind turbines are located to the east, continuously idling, and the west side features are mainly open shrubbery and grassland. The surface characteristics mainly consist of shrubs and grasslands. To the east, there is a secondary road connected to the main road, which may cause a certain degree of sound wave reflections. The three abandoned small wind turbines to the east may introduce interference in noise measurements, and when wind speeds are too high, they may adversely affect the measurement accuracy.

\section{Test Instruments}

\begin{figure}[H]
	\centering 
	\subfigure[Microphone on a mounting board]{ 
		\label{microphone}
		\includegraphics[height=.35\textwidth,trim=7 7 7 7,clip]{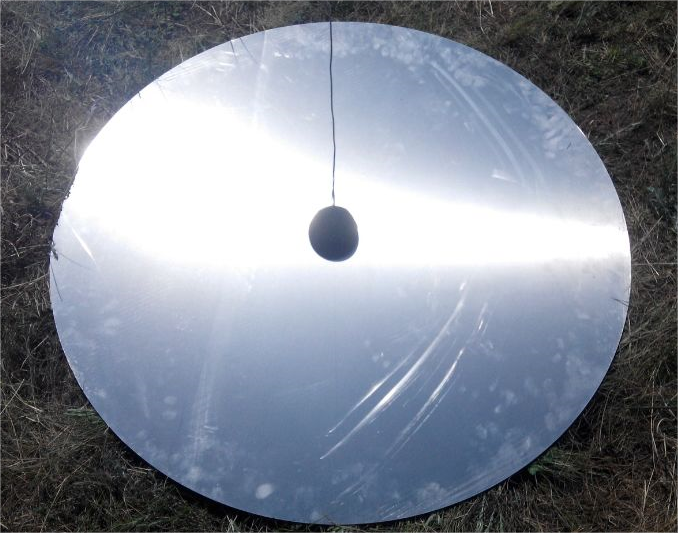} 
	} 
	\subfigure[NI digital acquisition system]{ 
		\label{NI}
		\includegraphics[height=.35\textwidth,trim=7 7 7 7,clip]{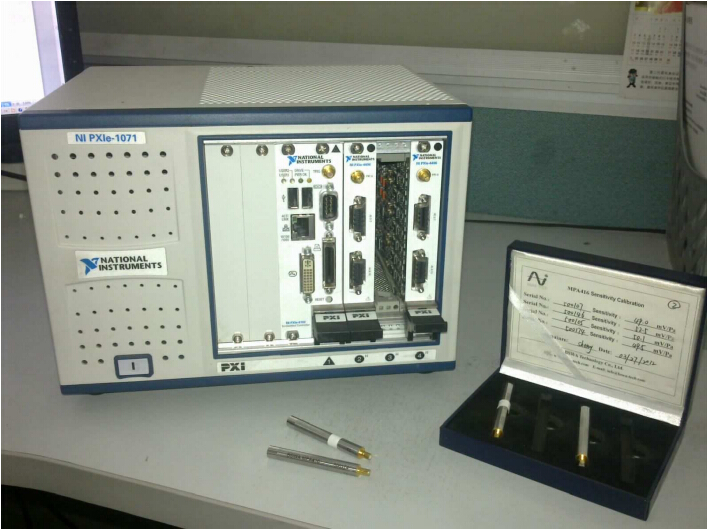} 
	} 
	\caption{Some main instruments} 
	\label{instruments}
\end{figure}

\begin{figure}[H]
	\centering
	\includegraphics[width=0.8\textwidth]{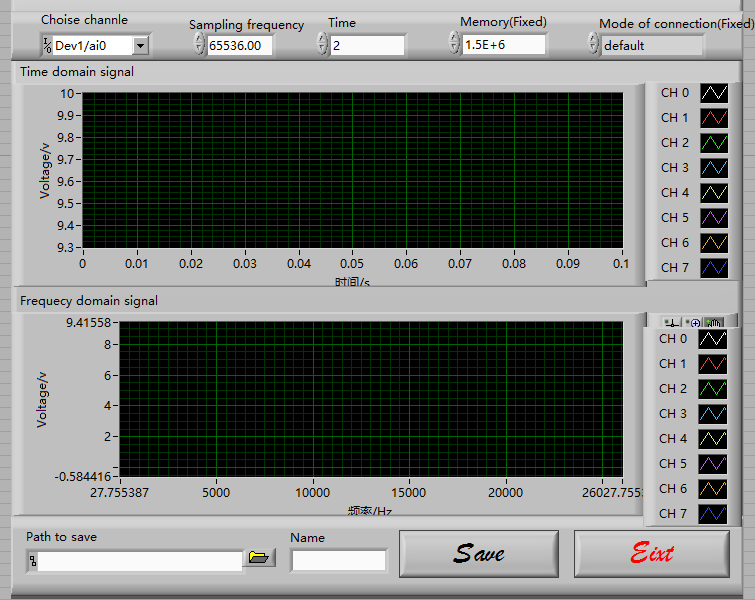}
	\caption{The Labview program for data acquisition and processing}
	\label{labview}
\end{figure}

\begin{enumerate}
\item A microphone with a baseboard and windscreen that meets the requirements of IEC 61400-11, as shown in Fig.~\ref{microphone}.
\item NI digital acquisition system, with the PXIe-4496 data acquisition card offering a single-channel synchronous sampling frequency of 204.8kS/s, meeting the sampling frequency requirements for noise measurement in this experiment, as shown in Fig.~\ref{NI}.
\item Labview data acquisition and processing program, including time domain data sampling, FFT transform to the frequency domain, sound pressure level calculation modules, etc., shown in Fig.~\ref{labview}.
\item Other instruments, such as a monitor, laser rangefinder, and data cables.
\end{enumerate}

\section{Data Processing Procedure}

The purpose of data processing is to obtain 1/3-octave sound power spectrum and overall power spectrum using statistical methods. There are two averaging methods used in this analysis: arithmetic averaging for non-acoustic data and energy averaging for acoustic data.

For the uncertainty calculation of the 1/3-octave sound power spectrum and overall power spectrum, this work follows the IEC 61400-11 standard and divides it into Type A uncertainty and Type B uncertainty. Type A uncertainty is obtained from statistical methods using multiple repeated measurements, while Type B uncertainty is derived from similar situations based on available information and experience. Finally, the Type A and Type B uncertainties are combined to obtain the combined standard uncertainty. It should be noted that the uncertainty of the acoustic measurement instrument chain cannot be accurately obtained due to the lack of instrument calibration certificates. In this work, it is assumed to be 0.2 dB. The values for other uncertainties are shown in Table~\ref{wind_speed_uncertain} and Table~\ref{uncertain}.

\begin{table}[H]
	\caption{Type B wind speed uncertainty component relevant for apparent sound power spectra}
	\centering
	\begin{tabular}{cc}
		\hline
		Uncertainty component& Standard uncertainty, m/s\\
        \hline
		Wind speed& 0.28\\
		\hline
	\end{tabular}
	\label{wind_speed_uncertain}
\end{table}

\begin{table}[H]
	\caption{Type B uncertainty component relevant for apparent sound power spectra}
	\centering
	\begin{tabular}{cc}
		\hline
		Uncertainty component& Standard uncertainty, dB\\
        \hline
		Calibration& 0.2\\
        Instrument& 0.2\\
        Borad& 0.3\\
        Wind screen insertion loss& 0\\
        Distance and direction& 0.1\\
        Air absorption& 0\\
        Weather conditions& 0.46\\
		\hline
	\end{tabular}
	\label{uncertain}
\end{table}

Noise and wind speed are measured for 10 seconds each time, and the data points are classified and averaged according to wind speed bins, resulting in:

\begin{enumerate}
    \item Average wind speed;
    \item Average A-weighted 1/3-octave spectrum;
    \item Corresponding standard uncertainty.
\end{enumerate}

For each 1/3-octave band, the background noise spectrum and total noise spectrum at the bin center can be obtained by linear interpolation of the average values of adjacent bins. At each wind speed bin center, the total noise spectrum is corrected using the 1/3-octave background noise spectrum at the same wind speed bin center, resulting in the 1/3-octave spectrum of the wind turbine generator. If the difference between the sum of the 1/3-octave band spectrum of the total noise and the sum of the 1/3-octave band spectrum of the background noise falls within the range of 3 to 6 dB, this result is marked with an asterisk (*) in the report. If this difference is less than or equal to 3 dB, the result corresponding to that wind speed bin is not recorded.

\begin{figure}[H]
	\centering
	\includegraphics[width=0.85\textwidth]{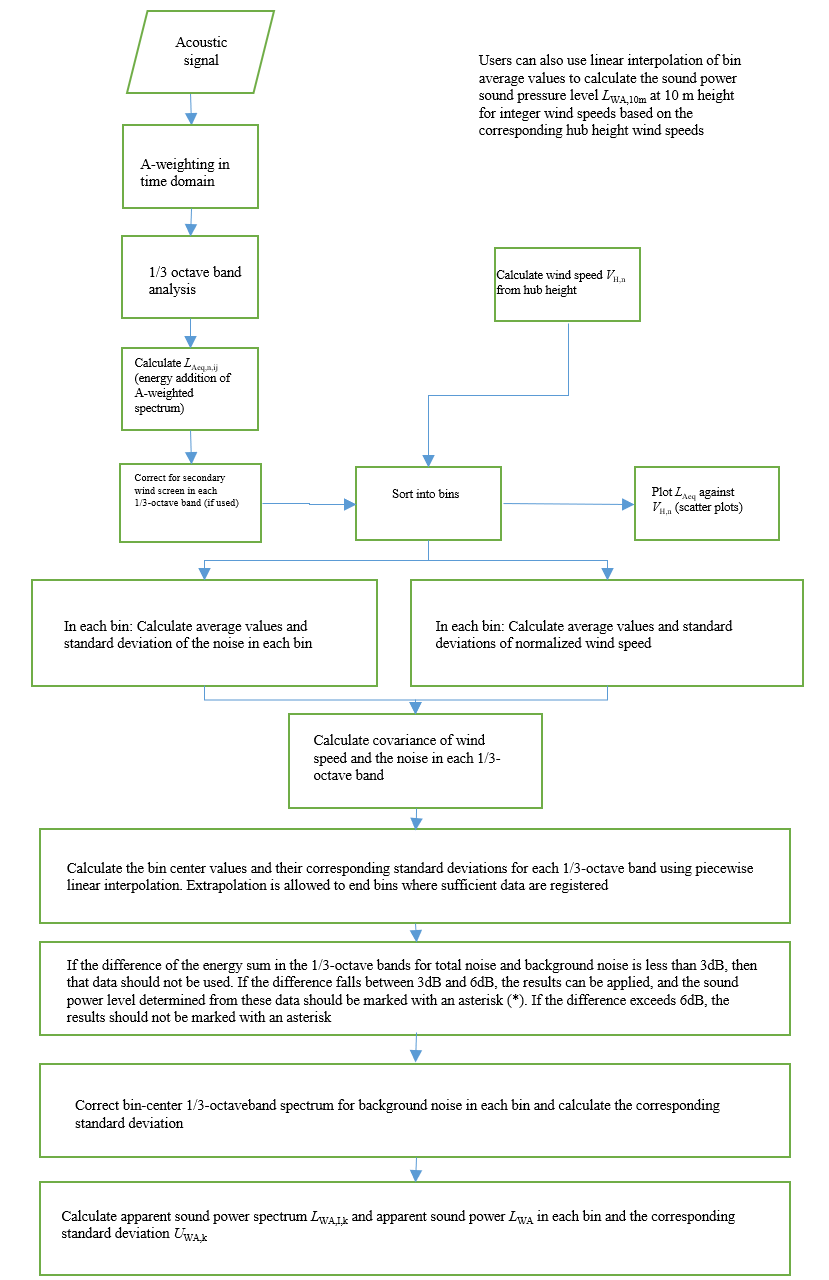}
	\caption{Flowchart showing the data processing procedure}
	\label{flowchart}
\end{figure}

The details of the data processing procedure are further described in the form of a flowchart~\ref{flowchart}.

\section{Sound Power Level}

It is important to select reasonable experimental data (discarding outliers), and make sure that noise and wind speed are classified into wind speed bins and averaged over a period of 10 seconds. The apparent sound power level and its uncertainty are calculated based on the reference wind speed at hub height, as shown in Table~\ref{apparent_SPL}. Alternatively, based on the experimental conditions in this paper, a reasonable roughness length $z_0$ = 0.05 m can be chosen, and the data processing can be carried out based on the wind speed at a reference height of 10 m using the logarithmic wind profile formula. Unless otherwise specified, the intermediate and final results analyzed and explained in this paper are based on the wind speed at hub height and will not be reiterated.

\begin{table}[H]
	\caption{Apparent sound power level (SPL)}
	\centering
	\begin{tabular}{ccc}
		\hline
		Hub height wind speed, m/s& Apparent SPL, dB& Uncertainty, dB\\
        \hline
		6& 96.65& 1.58\\
        7& 96.47& 1.30\\
        8& 97.02& 1.02\\
        9& 99.16& 0.94\\   
        10*& 98.40*& 1.74*\\        
		\hline
	\end{tabular}
	\label{apparent_SPL}
\end{table}

\begin{figure}[H]
	\centering
	\includegraphics[width=0.6\textwidth]{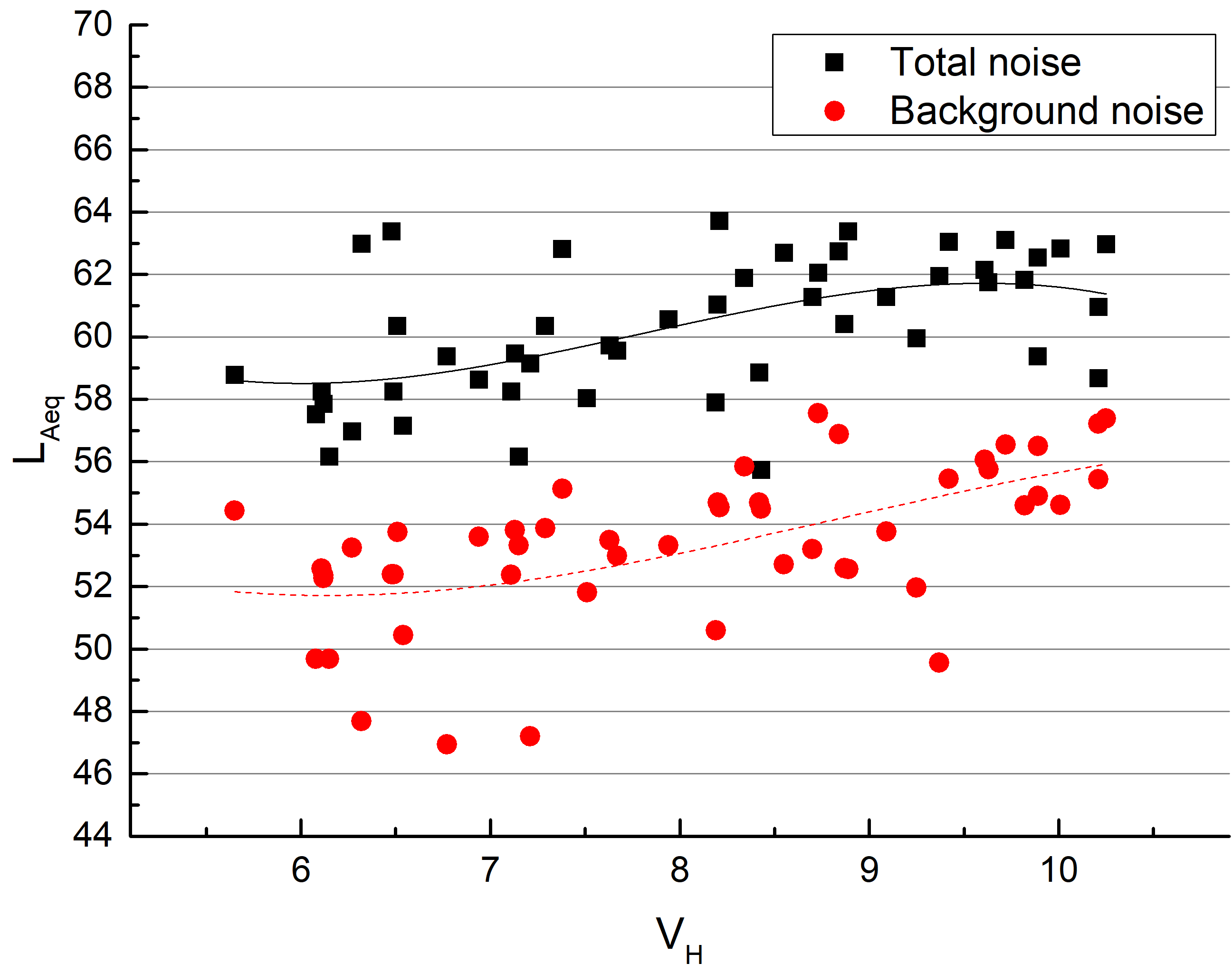}
	\caption{Scatter plot of total noise and background noise (A-weighted)}
	\label{scatter}
\end{figure}

From Fig.~\ref{scatter}, it can be observed that the differences between the sound pressure level of total noise and the sound pressure level of background noise for the wind speed bins of 6 m/s, 7 m/s, 8 m/s, and 9 m/s are all greater than 6 dB. For the 10 m/s bin, the difference falls within the range of [3, 6] dB, indicated by an asterisk in Table~\ref{apparent_SPL}. Within this range, the apparent sound power level shows an increasing trend with the increasing speed, and the rate of increase becomes faster. However, when the wind speed reaches 10 m/s, the apparent sound power level suddenly decreases. Analyzing the data for the 10 m/s bin marked with asterisks, it can be observed that the difference between the total noise and background noise decreases, and the more rapidly growing background noise has a negative impact on the results. Additionally, from the 6 m/s bin to the 7 m/s bin, the apparent sound power level decreases slightly, with values much smaller than the uncertainty. This may be related to changes in wind direction during the measurements and a slight increase in background noise.

Based on the analysis above, let's examine the relationship between the A-weighted sound pressure level of total noise and wind speed in the experiment, as shown in Fig.~\ref{scatter}. The fitted curve in Fig.~\ref{scatter} also indicates that there is a large difference between the total noise and background noise when the wind speed is low. As the wind speed increases to the bin of 10 m/s, the increase in the total noise slows down and even exhibits a downward trend. There is indeed a significant reduction in the difference between total noise and background noise. The reason for this might be a lack of secondary windshield, or increased variability in wind speed and wind direction under high wind speed conditions, leading to slightly unstable test conditions for the wind turbine's dynamic response.

\begin{figure}[H]
	\centering 
	\subfigure[Scatter plot of A-weighted sound pressure spectrum in the 1/3-octave band for the total noise]{ 
		\label{LAeqij}
		\includegraphics[width=.47\textwidth,trim=7 7 7 7,clip]{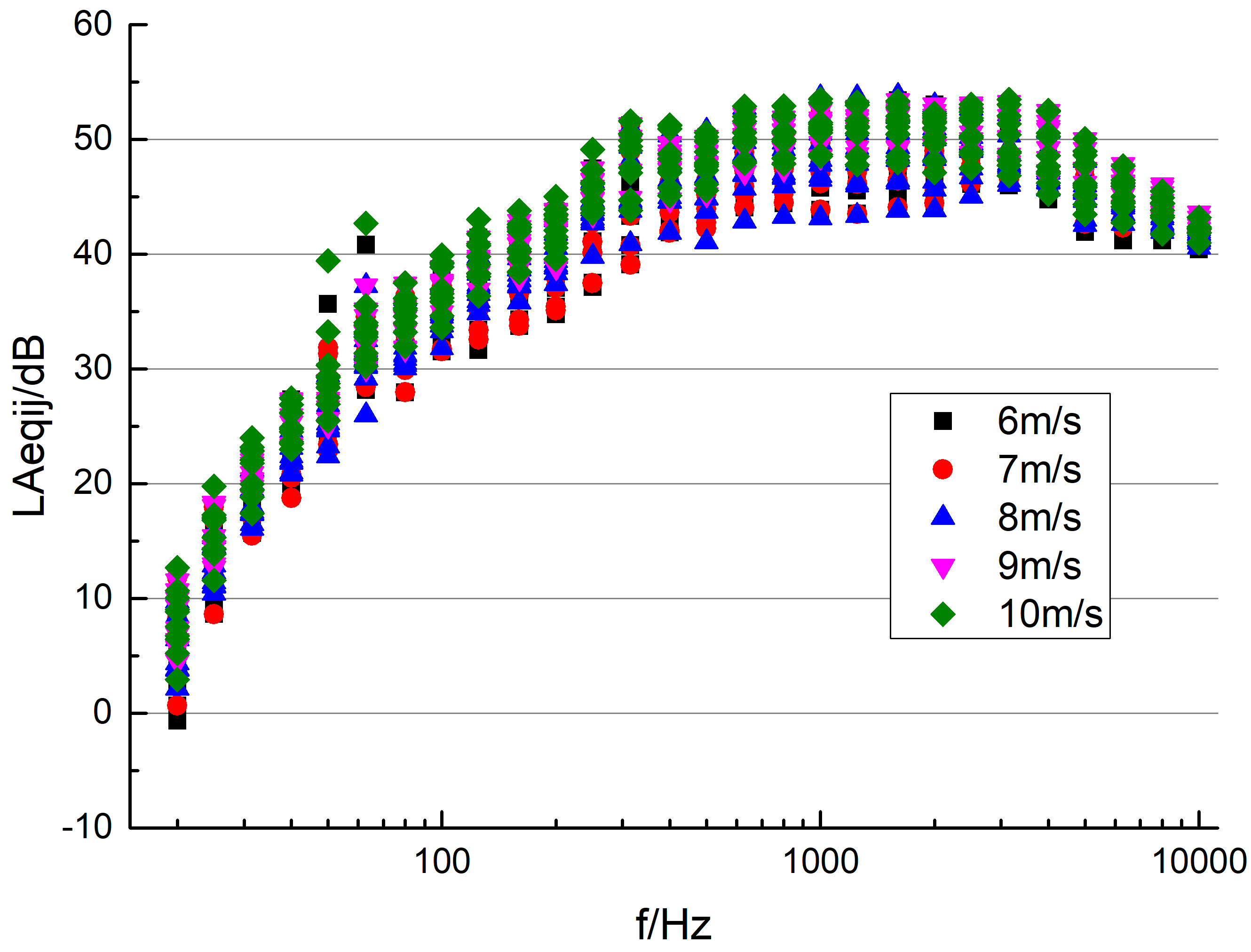} 
	} 
	\subfigure[Scatter plot of A-weighted sound pressure spectrum in the 1/3-octave band for the background noise]{ 
		\label{LAeqijB}
		\includegraphics[width=.47\textwidth,trim=7 7 7 7,clip]{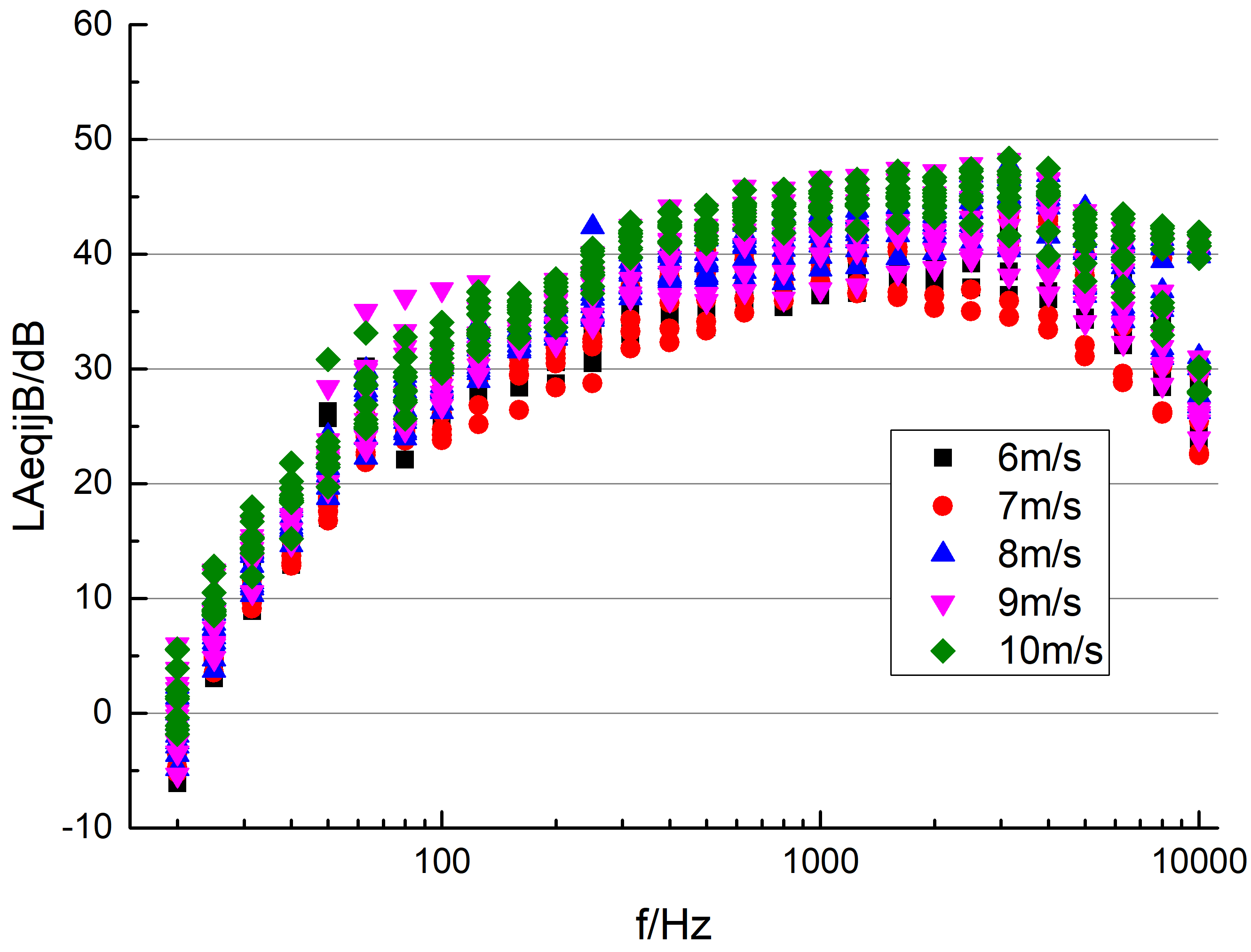} 
	} 
	\caption{Scattered data of A-weighted sound pressure spectra for the total noise and background noise} 
	\label{LAeq}
\end{figure}

According to the IEC 61400-11 wind turbine noise measurement standard, both total noise and background noise should be processed in a similar way. Fig.~\ref{LAeq} shows scatter plots of the A-weighted sound pressure spectrum in the 1/3-octave band of the total noise and background noise. Each wind speed bin contains 10 valid measurements. In general, as the wind speed increases, both the total noise and background noise increase, but this trend is not always consistent across all wind speed bins and frequency bands.

\begin{figure}[H]
	\centering 
	\subfigure[A-weighted sound pressure spectrum in the 1/3-octave band for the total noise]{ 
		\label{LVi}
		\includegraphics[width=.47\textwidth,trim=7 7 7 7,clip]{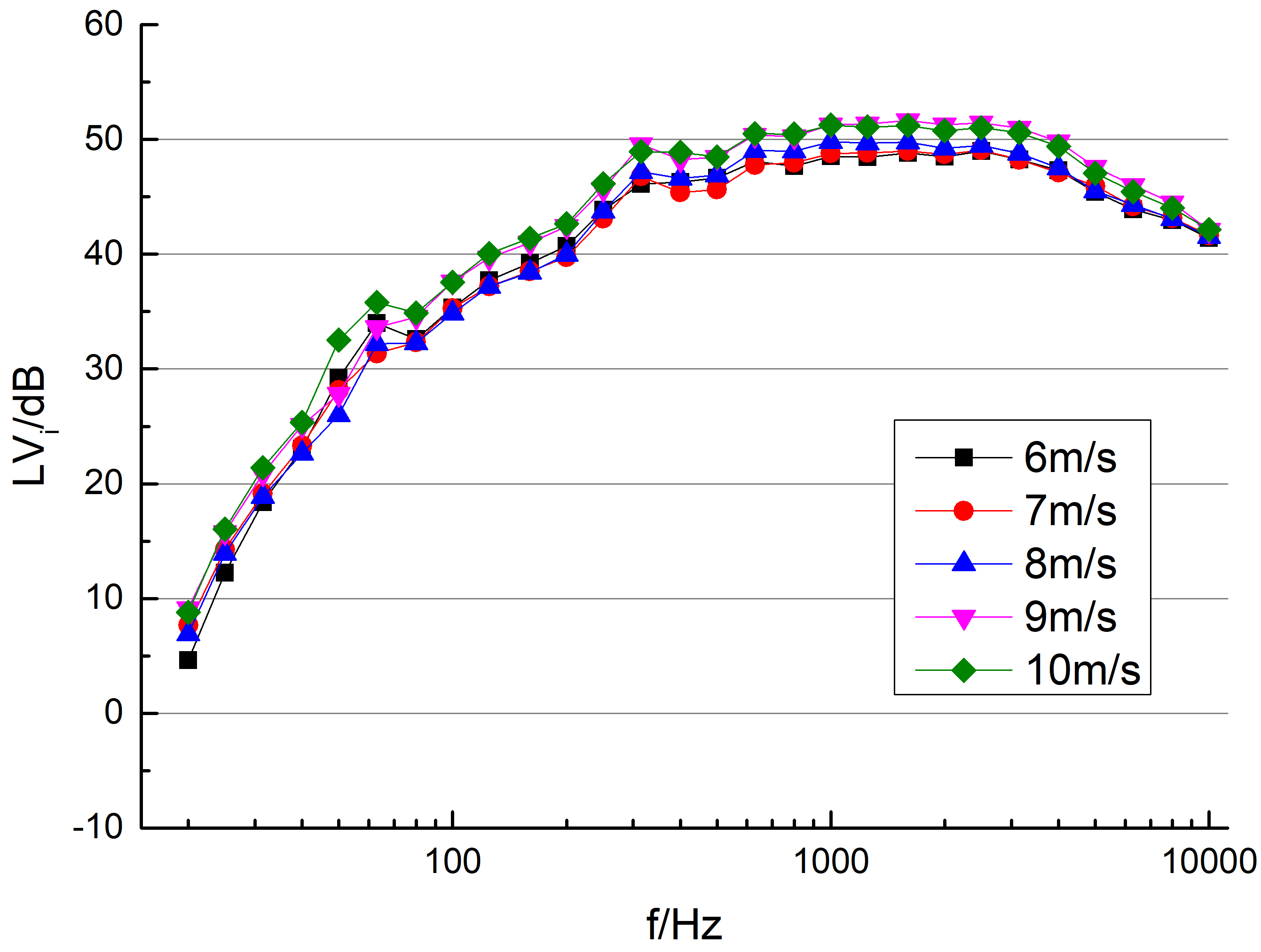} 
	} 
	\subfigure[A-weighted sound pressure spectrum in the 1/3-octave band for the background noise]{ 
		\label{LViB}
		\includegraphics[width=.47\textwidth,trim=7 7 7 7,clip]{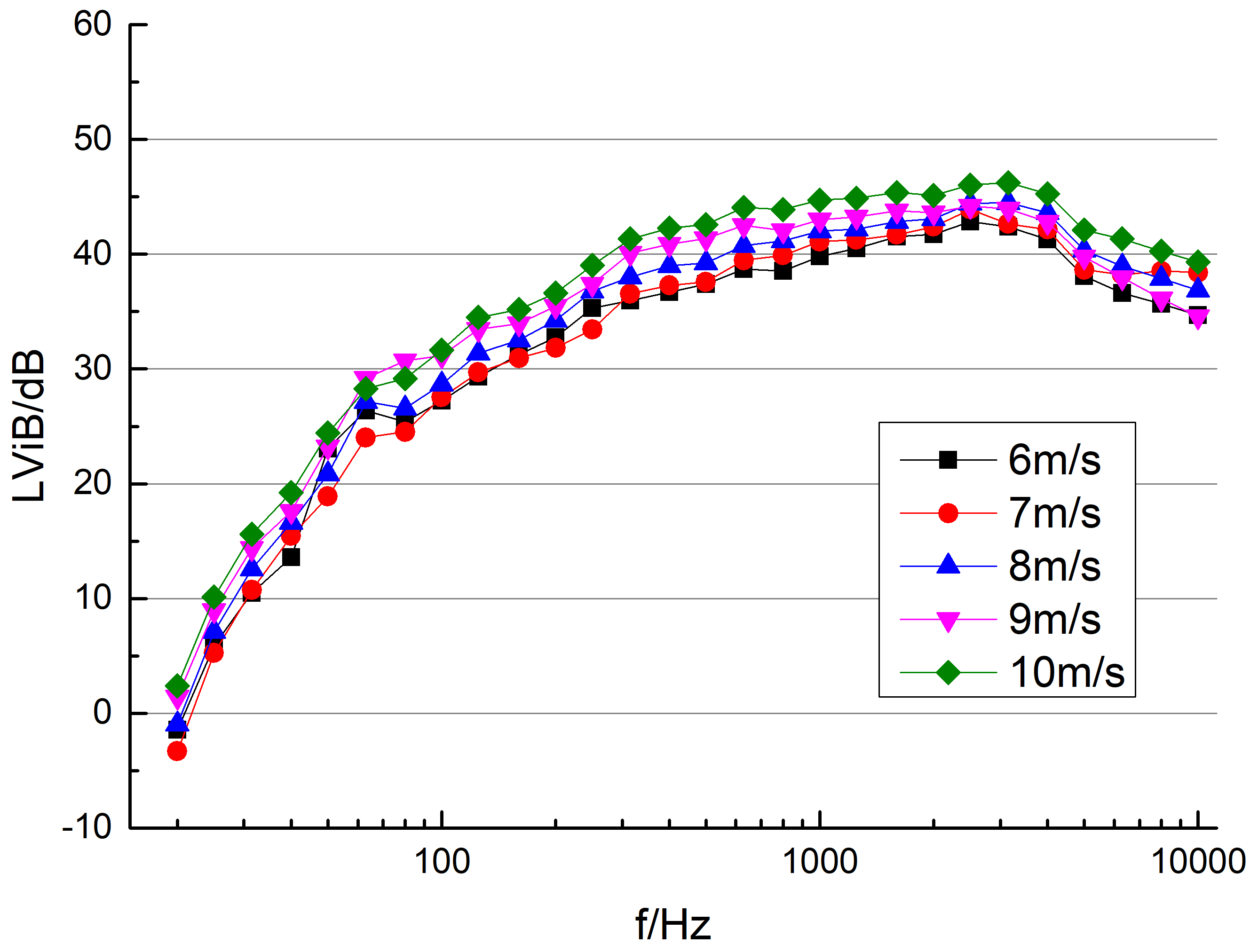} 
	} 
	\caption{A-weighted sound pressure spectra in the 1/3-octave band for the total noise and background noise} 
	\label{LVi_total}
\end{figure}

The A-weighted sound pressure spectra for the total noise and background noise from the 10 measurements are averaged over 1/3-octave frequency bands. The total noise and background noise at the center of each wind speed bin are obtained through linear interpolation of the average values of adjacent bins, as shown in Fig.~\ref{LVi_total}. Statistically, within the same frequency band, higher wind speeds correspond to higher total noise and background noise levels. Additionally, small peaks can be observed around f = 60Hz and f = 315Hz.

\begin{figure}[H]
	\centering 
	\subfigure[Standard uncertainty of the sound pressure spectrum for the wind speed bin centers of total noise]{ 
		\label{uLv}
		\includegraphics[width=.47\textwidth,trim=7 7 7 7,clip]{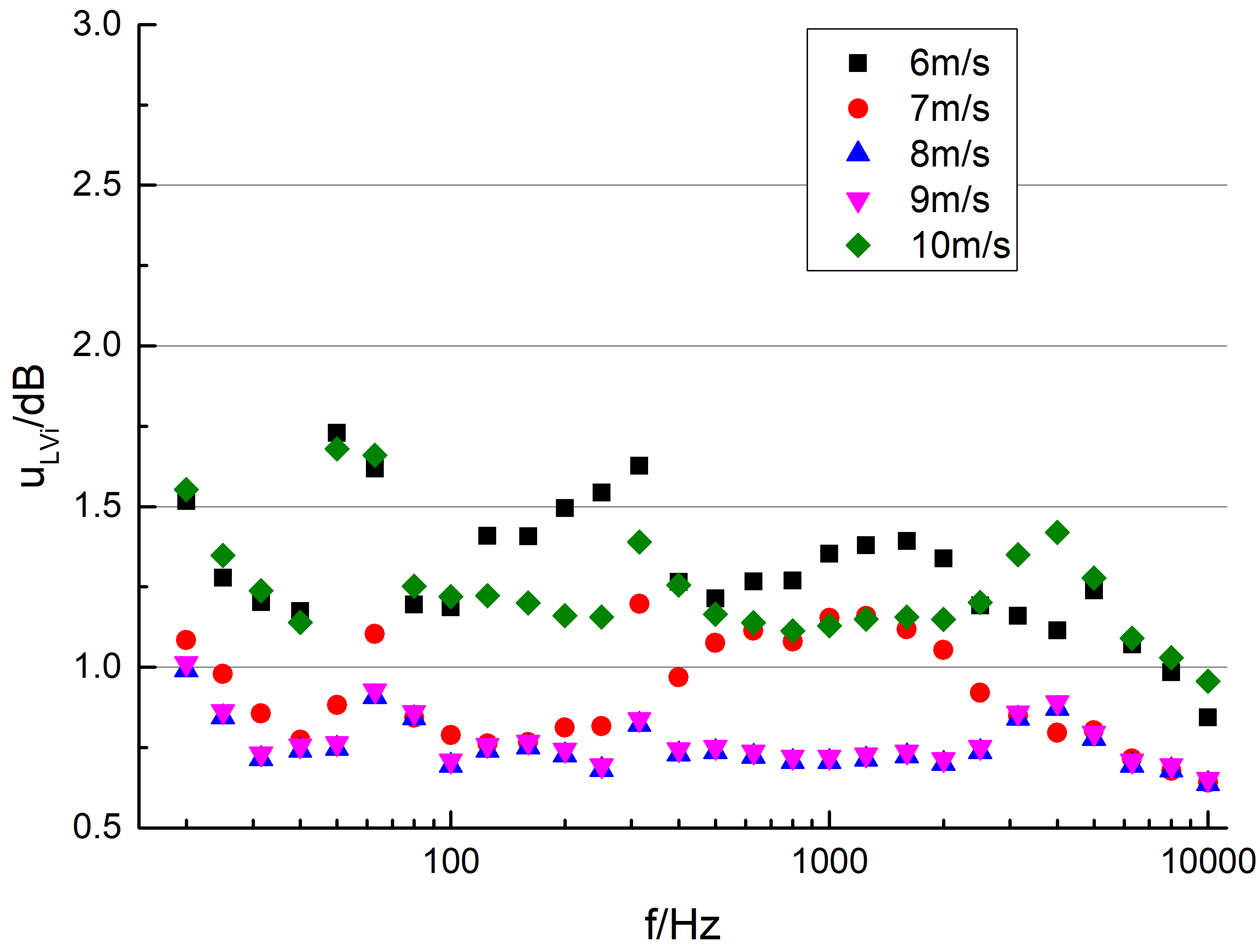} 
	} 
	\subfigure[Standard uncertainty of the sound pressure spectrum for the wind speed bin centers of background noise]{ 
		\label{uLvB}
		\includegraphics[width=.47\textwidth,trim=7 7 7 7,clip]{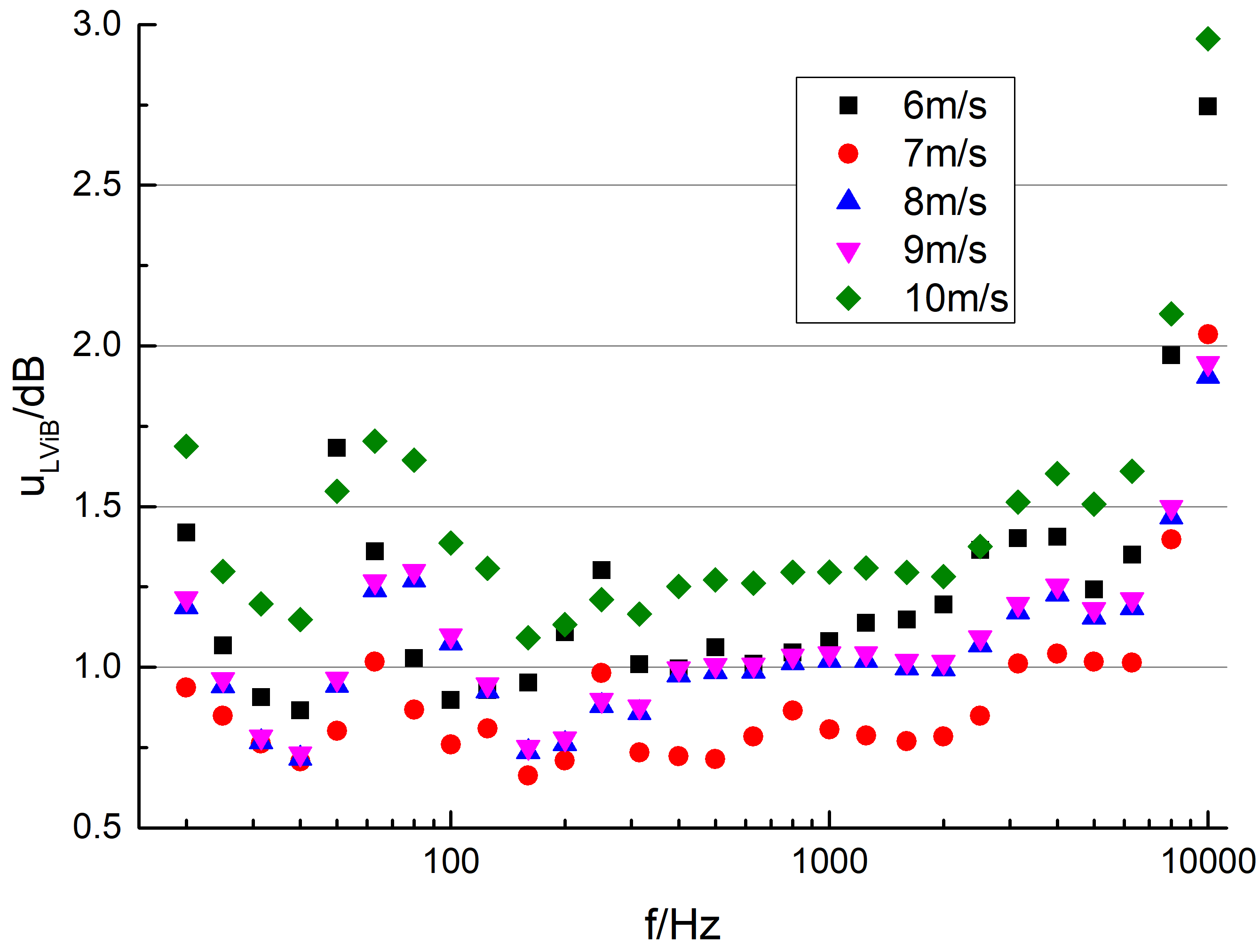} 
	} 
	\caption{Standard uncertainty of the sound pressure spectra of total noise and background noise} 
	\label{uncertainty}
\end{figure}

Fig.~\ref{uncertainty} shows the standard uncertainties of the sound pressure spectrum for the wind speed bin centers of total noise and background noise. The calculation of these uncertainties involves combining Type A and Type B uncertainties, incorporating covariance with wind speed, and substituting them into the formula for the standard uncertainty of the sound pressure spectrum at the bin center obtained through segmented linear interpolation. From Fig.~\ref{uncertainty}, it can be observed that the standard uncertainties of the sound pressure spectra for both the total noise and background noise are larger at higher or lower wind speeds. The uncertainties for the 8 m/s and 9 m/s wind speed bins are the smallest. This indicates that relatively large measurement errors occur at lower and higher wind speeds.

\begin{figure}[H]
	\centering 
	\subfigure[A-weighted corrected sound pressure spectrum in the 1/3-octave band]{ 
		\label{LVci}
		\includegraphics[width=.47\textwidth,trim=7 7 7 7,clip]{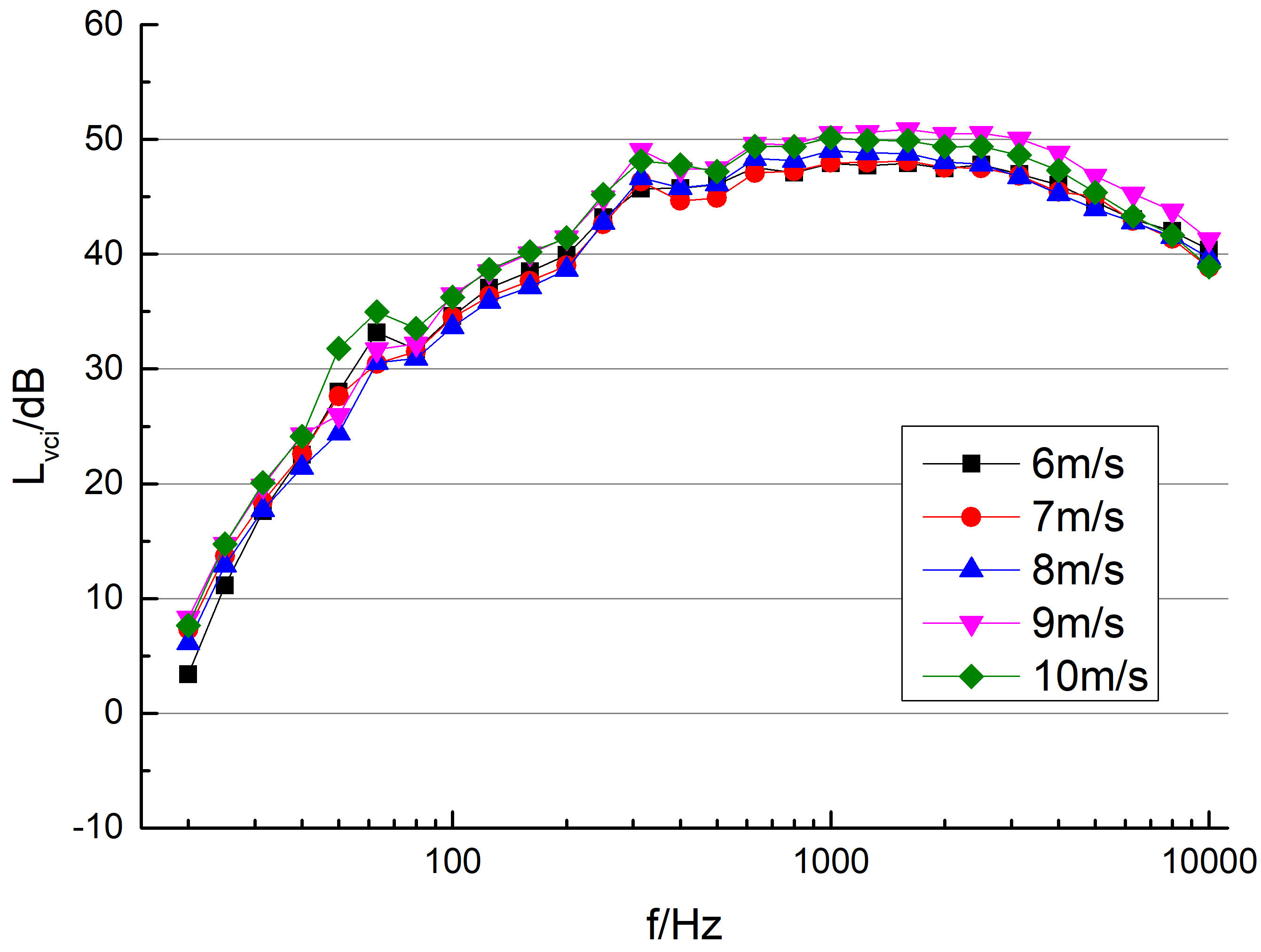} 
	} 
	\subfigure[Uncertainty of A-weighted corrected sound pressure spectrum in the 1/3-octave band]{ 
		\label{uLVci}
		\includegraphics[width=.47\textwidth,trim=7 7 7 7,clip]{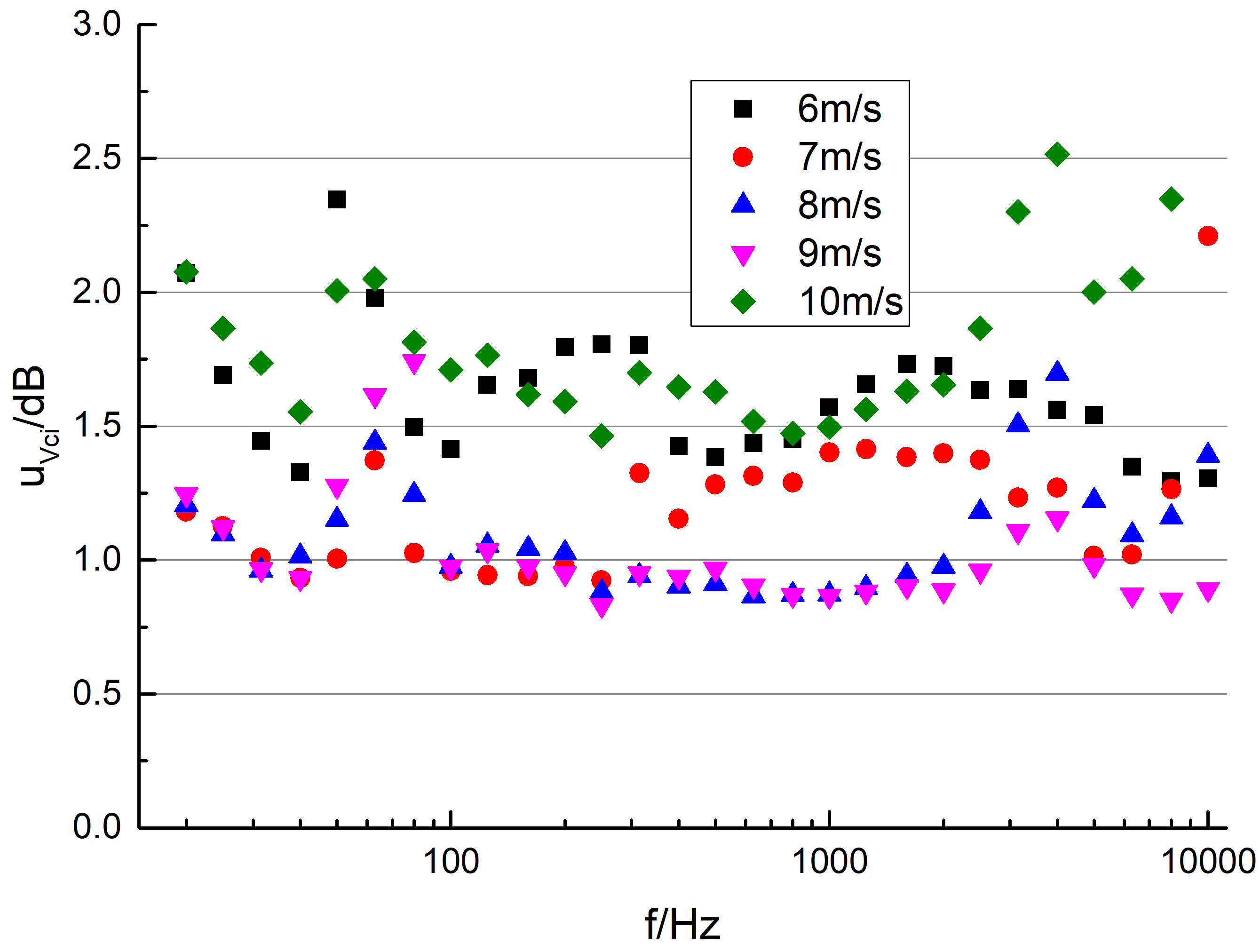} 
	} 
	\caption{Corrected A-weighted sound pressure spectrum and its uncertainty in the 1/3-octave band of total noise} 
	\label{corrected}
\end{figure}

\begin{figure}[H]
	\centering
	\includegraphics[width=0.6\textwidth]{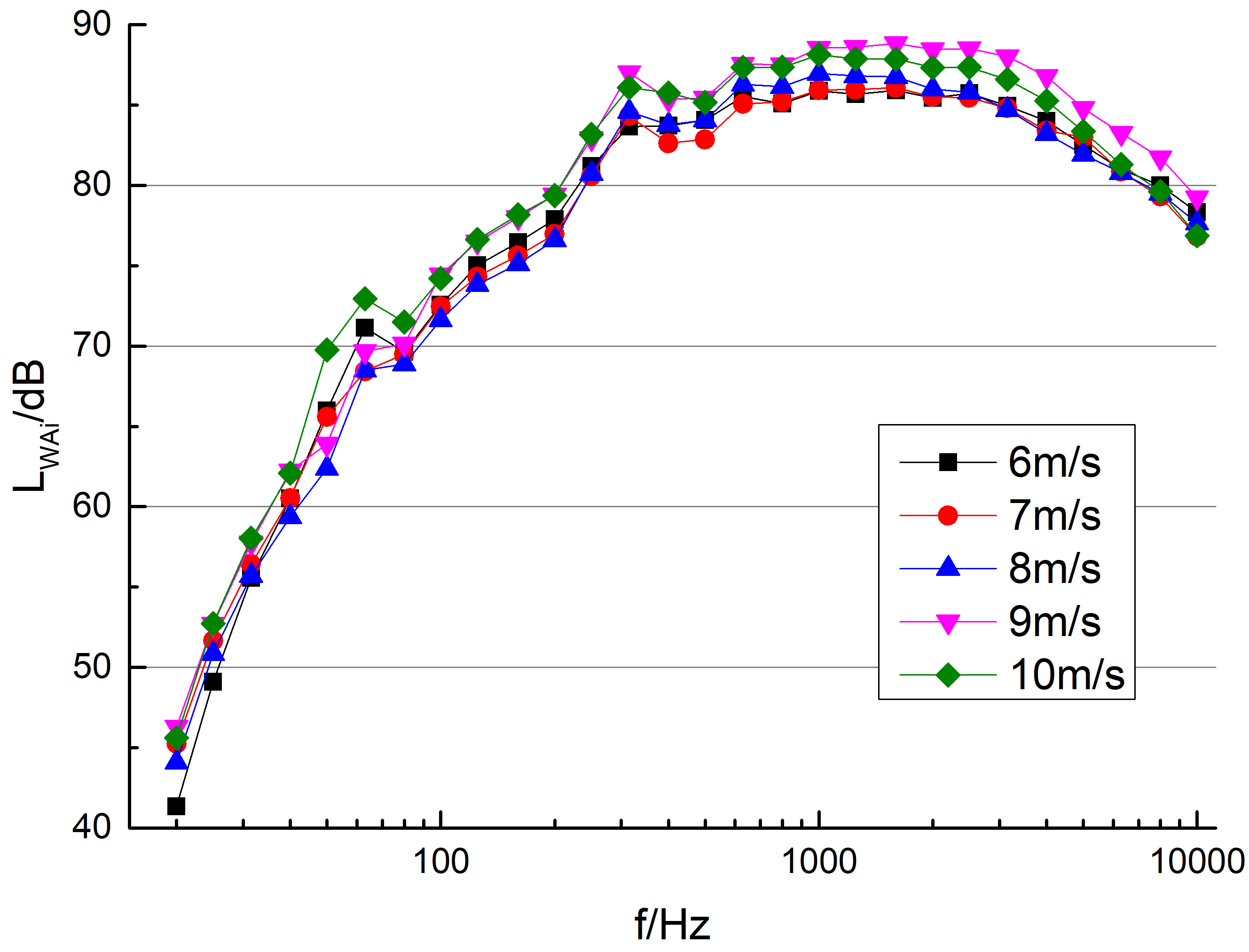}
	\caption{Apparent sound power spectrum in the 1/3-octave band}
	\label{apparent}
\end{figure}

Fig.~\ref{corrected} shows the A-weighted corrected sound pressure spectrum and its uncertainties in the 1/3-octave band, obtained by correcting the total noise based on the corresponding background noise. Similar to the previous analysis, small peaks can be observed around f = 60Hz and f = 315 Hz. Unlike the previous analysis, at lower frequencies, the corrected sound pressure level for higher wind speeds is higher than that for lower wind speeds. However, as the frequency increases, the highest corrected sound pressure level does not necessarily correspond to the highest wind speed. In terms of uncertainties, the highest uncertainties occur at both lower and higher wind speeds, while the uncertainties are smaller for wind speeds of 8 m/s and 9 m/s. Fig.\ref{apparent} represents the apparent sound pressure spectrum in the 1/3-octave band, which exhibits similar characteristics to Fig.~\ref{LVci}, but will not be analyzed further.

\section{Conclusions}

This study conducted noise measurements on a small-scale 100 kW wind turbine with large thickness and blunt trailing edge blades, following the IEC 61400-11 wind turbine noise measurement method. The characteristics of the wind turbine and the measurement environment were provided to facilitate future discussions on noise measurements of different wind turbines. The acoustic and non-acoustic data were analyzed, and the apparent sound power level referenced to the hub height wind speed was obtained. Furthermore, the intermediate results and their uncertainties were calculated and presented. Based on the experimental results presented in this work, the following conclusions can be drawn:

(1) Within the range of 6 - 9 m/s, the apparent sound power level shows an increasing trend with increasing wind speed, which is consistent with the characteristics of wind turbine noise. However, the rate of increase becomes faster and the apparent sound power level suddenly decreases when the wind speed reaches 10 m/s. The reason for this phenomenon may be that the total noise increases slowly or even decreases after the 10 m/s wind speed bin, causing the difference between total noise and background noise decreases.

(2) In the frequency bands of f = 60 Hz and f = 315 Hz, local maximas were observed. These two frequencies are suspicious "tone" frequencies and require additional valid experimental data.

(3) For both higher and lower wind speeds, the standard uncertainties of the sound pressure levels for the total noise and background noise are larger. This may be due to the relatively rapid changes in wind speed and wind direction during the measurement period, leading to increased dynamic characteristics of the wind turbine and unstable testing conditions.



%%       \citet{<label>} ==> Jones et al. [21]
%%       \citep{<label>} ==> [21]
%%

\bibliographystyle{elsarticle-num-names} 


%\begin{thebibliography}{00}



%\end{thebibliography}

\bibliography{mybib}

\end{document}